\documentstyle[sprocl,epsf]{article}

\def\be{\begin{equation}}
\def\ee{\end{equation}}
\def\bea{\begin{eqnarray}}
\def\eea{\end{eqnarray}}

\begin{document}
\thispagestyle{empty}
\setcounter{page}{0}
\begin{flushright}
SLAC-PUB-7501\\
BIHEP-TH-97-06
\end{flushright}

\vfill
\bigskip
\centerline{{\large \bf The Spin and Flavor Content
of Intrinsic Sea Quarks}\footnote{Work supported in part by
the Department of Energy, contract number DE--AC03--76SF00515.}}
\vspace{42pt}

\centerline{\bf Bo-Qiang Ma}
\vspace*{8pt}
\centerline{Institute of High Energy Physics, Academia Sinica}
\centerline{P.~O.~Box 918(4), Beijing 100039, China}
\vspace*{8pt}
\smallskip
\centerline{\bf Stanley J.~Brodsky}
\vspace*{8pt}\centerline{Stanford Linear Accelerator, Stanford University}
\centerline{Stanford, CA 94309, USA}

\bigskip

\vfill
\begin{center} {\large \bf Abstract}
\end{center}
The intrinsic quark-antiquark pairs generated by the
minimal energy
nonperturbative meson-baryon fluctuations in the nucleon sea provide a
consistent framework for understanding a number of empirical
anomalies observed in the deep inelastic
quark-parton structure of nucleons:
the flavor asymmetry of the
nucleon sea implied by the violation of Gottfried sum rule,
the proton spin problem implied by
the violation of the Ellis-Jaffe sum rule,
and the outstanding conflict between two different determinations of the
strange quark sea in the nucleon.

\bigskip
\centerline{Invited talk at the International Conference on
Perspective in Hadronic Physics}
\smallskip
\centerline{12-16 May 1997, Trieste, Italy}

\vfill
\newpage
\title{The Spin and Flavor Content
of Intrinsic Sea Quarks
}

\author{BO-QIANG MA}

\address{Institute of High Energy Physics, Academia Sinica\\
P.~O.~Box 918(4), Beijing 100039, China}

\author{STANLEY J.~BRODSKY}

\address{Stanford Linear Accelerator, Stanford University\\ 
Stanford, CA 94309, USA}


\maketitle
\abstracts{
The intrinsic quark-antiquark pairs generated by the
minimal energy
nonperturbative meson-baryon fluctuations in the nucleon sea provide a
consistent framework for understanding a number of empirical
anomalies observed in the deep inelastic
quark-parton structure of nucleons:
the flavor asymmetry of the
nucleon sea implied by the violation of Gottfried sum rule,
the proton spin problem implied by
the violation of the Ellis-Jaffe sum rule,
and the outstanding conflict between two different determinations of the
strange quark sea in the nucleon.
}
  
\section{Introduction}

By far the most unexpected empirical features of the structure of
hadrons relate to the composition of the nucleons
in terms of their nonvalence
quarks.  For example, the violation of the Gottfried  sum rule measured by
the New Muon Collaboration
(NMC) 
indicates a strong violation of SU(2) symmetry in the
$\bar u $ and $\bar d$ distributions \cite{Kum97}.
The large violation of the Ellis-Jaffe sum rule as
observed at CERN 
and SLAC 
indicates  
that only a small fraction of the proton's helicity is provided by
quarks \cite{SpinR}.
The European Muon Collaboration (EMC) has
observed a large excess of charm quarks at large momentum fraction $x$ in
comparison with the charm distributions predicted from photon-gluon fusion
processes \cite{EMC82}.  Furthermore, as we have recently
emphasized \cite{Bro96}, it
is not even clear that the quark and antiquark sea distributions
have identical
momentum and helicity
distributions, contrary to intuition based on perturbative gluon-splitting
processes.

It is important to distinguish two distinct types of quark and
gluon contributions to the nucleon sea measured in deep inelastic
scattering: ``extrinsic" and ``intrinsic". 
We shall refer to the sea quarks generated from
the QCD hard bremsstrahlung and gluon-splitting
as ``extrinsic" quarks, since the sea quark structure
is associated with the internal composition of gluons, rather than the
proton itself.  In contrast, sea quarks which are multi-connected to the
valence quarks of the nucleon are referred to as ``intrinsic" sea quarks
\cite{Bro96,Bro81}.
In this talk we shall show that the intrinsic quark-antiquark pairs generated
by the
minimal energy
nonperturbative meson-baryon fluctuations in the nucleon sea provide a
consistent framework for understanding the origin of polarized
light-flavor and strange sea quarks
implied by the violation of the Ellis-Jaffe sum rule.
Furthermore,
the meson-baryon fluctuations of the
nucleon sea cause striking quark/antiquark asymmetries in the momentum and
helicity distributions for the down and strange contributions to the proton
structure function: the intrinsic
$d$ and $s$ quarks in the proton sea are predicted to be negatively polarized,
whereas the intrinsic $\bar d$ and $\bar s$
antiquarks give zero contributions
to the proton spin. 
The momentum distribution
asymmetry for strange quarks and antiquarks is supported by
an
outstanding conflict between two
different determinations of the strange quark
sea in the nucleon. The model predicts an excess of intrinsic $d
\bar d$ pairs over $u \bar u$ pairs, as supported by the Gottfried sum rule
violation.

In Fock state wavefunctions containing heavy quarks, the minimal energy
configuration occurs when the constituents have similar rapidities. 
Thus one of
the most natural features of intrinsic heavy sea quarks is their contribution to
the nucleon structure functions at large
$x$ in contrast to the small $x$ heavy quark distributions predicted from
photon-gluon fusion processes. 
This feature of intrinsic charm is supported by
the EMC
observation of a large excess of charm quarks at large $x$
\cite{EMC82,Bro81,Vog96}.

\section{The meson-baryon fluctuation model of intrinsic
$q \bar q$ pairs}

The intrinsic sea quarks and gluons are multi-connected to
the valence quarks and exist over a relatively long lifetime within
the nucleon bound state. Thus the intrinsic $q \bar q$ pairs can
arrange themselves together with the valence quarks of the target
nucleon into the most energetically-favored meson-baryon
fluctuations.
For example,
the coupling of a proton to a virtual $K^+ \Lambda$ pair provides a
specific source of intrinsic strange quarks and antiquarks in the
proton.  Since the $s$ and $\bar s$ quarks appear in different
configurations in the lowest-lying hadronic pair states, their
helicity and momentum distributions are distinct.  In our analysis,
we have utilized \cite{Bro96} a
boost-invariant light-cone Fock state description of the hadron
wavefunction which emphasizes multi-parton configurations of
minimal invariant mass.

In order to characterize the momentum and helicity distributions of
intrinsic $q \bar q$ pairs, we adopt a light-cone two-level
convolution model of structure functions \cite{Ma91} in which the
nucleon is a two-body system of meson and baryon which are also
composite systems of quarks and gluons. 
The intrinsic strangeness fluctuations in the
proton wavefunction are mainly due to the intermediate $K^+ \Lambda$
configuration since this state has the lowest off-shell light-cone
energy and invariant mass. The $K^+$ meson is a
pseudoscalar particle with negative parity, and the $\Lambda$ baryon
has the same parity of the nucleon. 
We thus write the
total angular momentum space wavefunction of the intermediate $K
\Lambda$ state in the center-of-mass reference frame as
\begin{eqnarray}
\left|J=\frac{1}{2},J_z=\frac{1}{2}\right\rangle
&=&\sqrt{\frac{2}{3}} \left|L=1,L_z=1\right\rangle\
\left|S=\frac{1}{2},S_z=-\frac{1}{2}\right\rangle\nonumber\\
&-&\sqrt{\frac{1}{3}}\
\left|L=1,L_z=0\right\rangle \
\left|S=\frac{1}{2},S_z=\frac{1}{2}\right\rangle\ .
\end{eqnarray}
Thus the
intrinsic strange quark normalized to the probability
$P_{K^+\Lambda}$ of the $K^+\Lambda$ configuration yields a
fractional contribution $\Delta S_{s}=2
S_z(\Lambda)=-\frac{1}{3}P_{K^+\Lambda} $ to the proton spin,
whereas the intrinsic antistrange quark gives a zero contribution:
$\Delta S_{\bar s}=0.$ There thus can be a significant quark and
antiquark asymmetry in the quark spin distributions for the
intrinsic $s \bar s$ pairs.

The quark helicity projections measured in deep inelastic scattering
are related to the quark spin projections in the target rest frame
by multiplying by a Wigner rotation factor of order 0.75 for light
quarks and of order 1 for heavy quarks \cite{Ma91b}.  We therefore
predict that the net strange quark helicity arising from the
intrinsic $s \bar s$ pairs in the nucleon wavefunction is negative,
whereas the net antistrange quark helicity is approximately zero.

\begin{figure}[htb] 
\begin{center}
\mbox{\epsfysize=8.0cm \epsffile{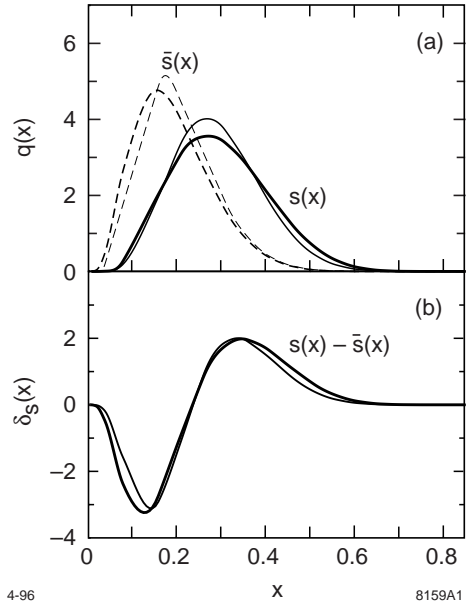}}
\end{center}
\caption[*]{\baselineskip 8pt  
The momentum distributions for the
strange quarks and antiquarks in the light-cone meson-baryon
fluctuation model of intrinsic $q \bar q$ pairs, with the
fluctuation wavefunction of $K^+\Lambda$ normalized to 1. The curves
in (a) are the calculated results of $s(x)$ (solid curves) and $\bar
s(x)$ (broken curves) with the Gaussian type (thick curves) and
power-law type (thin curves) wavefunctions and the curves in (b) are
the corresponding $\delta_s(x)=s(x)-\bar s(x)$. ~~~~~ 
}
\label{bmfig1}
\end{figure}

The momentum distributions of the intrinsic strange and antistrange
quarks in the $K^+ \Lambda$ state can be modeled from the two-level
convolution formula
\begin{equation}
s(x)=\int_{x}^{1} \frac{{\rm d}y}{y}
f_{\Lambda/K^+\Lambda}(y) q_{s/\Lambda}\left(\frac{x}{y}\right);
\;
\bar s(x)=\int_{x}^{1} \frac{{\rm d}y}{y}
f_{K^+/K^+\Lambda}(y) q_{\bar s/K^+}\left(\frac{x}{y}\right),
\end{equation}
where $f_{\Lambda/K^+\Lambda}(y)$, $f_{K^+/K^+\Lambda}(y)$ are
probability distributions of 
finding $\Lambda$, $K^+$ in the $K^+ \Lambda$ state
with the light-cone momentum fraction $y$ and $q_{s/\Lambda}(x/y)$,
$q_{\bar s/K^+}(x/y)$ are probability distributions of finding strange,
antistrange quarks in $\Lambda$, $K^+$ with the light-cone momentum
fraction $x/y$. We estimate these quantities by adopting
two-body momentum wavefunctions for $p=K^+ \Lambda$, $K^+=u {\bar
s}$, and $\Lambda=s u d$ where the $u d$ in $\Lambda$ serves as a
spectator in the quark-spectator model \cite{Ma96}. 
We calculated the
momentum distributions $s(x)$, $\bar s(x)$, and $\delta
_s(x)=s(x)-\bar s(x)$ and found
a significant quark/antiquark 
asymmetry of the momentum distributions for the strange sea quarks 
\cite{Bro96}, as shown in Fig.~\ref{bmfig1}.

We have performed similar calculations for the momentum
distributions of the intrinsic $d \bar d$ and $c \bar c$ pairs
arising from the $p(uud d \bar d )=\pi^{+}(u\bar{d})n(udd)$ and
$p(uud c \bar c )=\bar{D}^0(u\bar c) \Lambda^{+}_{c}(udc)$ 
configurations. 
The $c \bar{c}$
momentum asymmetry is small compared with the $s \bar{s}$ and $d
\bar{d}$ asymmetries but is still nontrivial. The $c \bar{c}$ spin
asymmetry, however, is large. Considering that it is difficult to
observe the momentum asymmetry for the $d \bar d$ pairs due to an
additional valence $d$ quark in the proton, the momentum asymmetry
of the intrinsic strange and antistrange quarks is the most
significant feature of the model and the easiest to observe.

\section{The light-flavor sea quark content and the Gottfried sum
rule violation}

Parton sum rules provide information on the quark distributions    
in nucleons
and thus allow for sensitive investigations of the detailed flavor and spin
content of nucleons. 
The Gottfried sum rule (GSR) 
violation reported
by the New Muon Collaboration (NMC) \cite{NMC91} has inspired a number of
discussions on  the flavor dependence of sea distributions in the
nucleons.
The Gottfried sum
is expressed as
\begin{equation}                        
S_{G}=\int_{0}^{1}[F_{2}^{p}(x_B)-F_{2}^{n}(x_B)]\frac{{\rm d} x_B}{x_B
}
=\frac{1}{3}+\int_{0}^{1}\sum_{i} [2\overline{q}_{i}^{p}(x_B)
 -2\overline{q}_{i}^{n}(x_B)]\;{\rm d} x_B.
\label{eq:ngsp}
\end{equation}
Under the assumptions of isospin symmetry between
proton and neutron, and flavor
symmetry in the sea, one arrives at the Gottfried sum
rule (GSR), $S_{G}=1/3$.
However, the value of $S_{G}$ measured in the NMC experiment is
\begin{equation}
S_{G}=0.235\pm0.026,
\end{equation}
which is significantly smaller than the simple quark-parton-model result of
1/3.
Several different explanations for the
origin of the GSR violation have been proposed,
such as flavor asymmetry of the nucleon sea \cite{Pre91,Pi},
isospin symmetry breaking between the proton
and the neutron \cite{Ma92} 
{\it et al.}.
Among the various explanations,
the most natural one is the $u$ and $d$ flavor asymmetry of the
nucleon sea which is implied by the meson cloud of the nucleon
\cite{Pi}.

The light-cone meson-baryon fluctuation model
contains neutral meson fluctuation
configurations in which the intermediate mesons are composite
systems of the intrinsic up $u \bar u$ and down $d \bar d$ pairs,
but these fluctuations do not cause a flavor asymmetry in
the nucleon sea. The lowest nonneutral $u \bar u$ fluctuation in
the proton is
$\pi^{-}(d \bar u)\Delta^{++}(uuu)$,  and its
probability is small compared to the less massive nonneutral $d \bar d$
fluctuation $\pi^{+}(u \bar{d})n(udd)$. 
Therefore the dominant nonneutral light-flavor $q \bar
q$ fluctuation in the proton sea is $d \bar d$ through the
meson-baryon configuration
$\pi^{+}(u \bar{d})n(udd)$. This
leads naturally to an excess of $d \bar d$ pairs over $u \bar u$
pairs in the proton sea.  Such a mechanism provides a natural
explanation \cite{Pi} for the violation of the Gottfried sum rule
\cite{NMC91} and leads to nontrivial distributions of the sea
quarks. The NMC measurement $ S_G=\frac{1}{3}+\frac{2}{3}
\int_0^1{\rm d} x [u_s(x)-d_s(x)] =0.235 \pm 0.026 $ \cite{NMC91}
implies $\int_0^1{\rm d} x [d_s(x)-u_s(x)]=0.148\pm 0.039$, which
can be considered as the probability of finding nonneutral
intrinsic $d \bar d$ fluctuations in the proton sea.

Thus, as a first support, the meson-baryon fluctuation model of intrinsic
$q \bar q$ pairs provides a natural
mechanism for the $u \bar u$ and $d \bar d$
asymmetry in the nucleon sea which is responsible for
the most part of the Gottfried sum rule violation.

\section{The spin content of the nucleon and the Ellis-Jaffe sum rule
violation}

The observation of the Ellis-Jaffe sum rule (EJSR) 
violation in
the inclusive polarized deep inelastic scattering experiments
has received an extensive attention
on the spin content of nucleons. The experimental data
of the integrated spin-dependent structure functions for the nucleons
are generally
understood to imply that the sum of the up, down, and strange
quark helicities in the nucleon is much smaller than the nucleon
spin.
There has been a number of possible interpretations
for the EJSR violation, and the
quark helicity distributions for
each flavor
are quite different
in these interpretations \cite{SpinR}. In most interpretations
the Ellis-Jaffe sum rule violation is considered to be independent
of the Gottfried sum rule violations. However, there have
been speculations and suggestions about the interrelation
between the two sum rule violations. It has been known
\cite{Bro96}
that the intrinsic
$q \bar q$ pairs generated by the nonperturbative
meson-baryon fluctuations in the nucleon sea, combined with the
flavor asymmetry in the valence component of the nucleon
\cite{Ma96} and
the Wigner rotation effect due to the quark relativistic transversal
motions \cite{Ma91b}, could provide a comprehensive picture to understand
a number of phenomena related to the proton spin problem
caused by the Ellis-Jaffe sum rule violation.

The unpolarized valence quark distributions $u_v(x)$ and $d_v(x)$ 
in the SU(6) quark-spectator model \cite{Ma96}
are expressed by
\begin{equation}
\begin{array}{clcr}
u_{v}(x)=\frac{1}{2}a_S(x)+\frac{1}{6}a_V(x);\\
d_{v}(x)=\frac{1}{3}a_V(x),
\label{eq:ud}
\end{array}
\end{equation}
where $a_D(x)$ ($D=S$ for scalar spectator or $V$ for vector
spectator) is normalized such
that $\int_0^1 d x a_D(x)=3$ and denotes the amplitude for the quark
$q$ is scattered while the spectator is in the diquark state $D$.
Exact SU(6) symmetry provides the relation $a_S(x)=a_V(x)$,
which implies the valence flavor symmetry $u_{v}(x)=2 d_{v}(x)$. This
gives the prediction $F^n_2(x)/F^p_2(x)\geq 2/3$ for
all $x$ 
which is ruled out by the experimental
observation $F^n_2(x)/F^p_2(x) <  1/2$ for $x \to 1$.
The mass
difference between the scalar
and vector spectators can reproduce the up and down valence
quark asymmetry that
accounts for the observed ratio $F_2^{n}(x)/F_2^{p}(x)$ at large $x$
\cite{Ma96}. 
This
supports the quark-spectator picture of deep inelastic scattering
in which the difference between the mass of the scalar and vector
spectators is important to reproduce the explicit
SU(6) symmetry breaking while the bulk SU(6) symmetry of the
quark model still holds.

The quantity $\Delta q$ measured in polarized deep inelastic
scattering is defined by the axial current matrix element
\begin{equation}
\Delta q=<p,\uparrow|\overline{q} \gamma^{+} \gamma_{5} q|p,\uparrow>.
\end{equation}
In the light-cone or quark-parton descriptions,
$\Delta q (x)=q^{\uparrow}(x)-q^{\downarrow}(x)$,
where $q^{\uparrow}(x)$ and $q^{\downarrow}(x)$ are the probability
distributions
of finding a quark or antiquark with longitudinal momentum
fraction $x$ and polarization parallel or antiparallel
to the proton helicity in the infinite momentum frame.
However, in the proton rest frame, one finds,
\begin{equation}
\Delta q (x)
=\int [{\rm d}^2{\bf k}_{\perp}] W_D(x,{\bf k}_{\perp})
[q_{s_z=\frac{1}{2}}
(x,{\bf k}_{\perp})-q_{s_z=-\frac{1}{2}}(x,{\bf k}_{\perp})],
\end{equation}
with
\begin{equation}
W_D(x,{\bf k}_{\perp})=\frac{(k^+ +m)^2-{\bf k}^2_{\perp}}
{(k^+ +m)^2+{\bf k}^2_{\perp}}
\end{equation}
being the contribution from the relativistic effect due to
the quark transversal motions \cite{Ma91b},
$q_{s_z=\frac{1}{2}}(x,{\bf k}_{\perp})$
and $q_{s_z=-\frac{1}{2}}(x,{\bf k}_{\perp})$ being the probability
distributions
of finding a quark and antiquark with rest mass $m$
and with spin parallel and anti-parallel to the rest proton
spin, and $k^+=x {\cal M}$ where
${\cal M}=\frac{m^2_q+{\bf k}^2_{\perp}}{x}
+\frac{m^2_D+{\bf k}^2_{\perp}}{1-x}$.
The Wigner rotation factor $W_D(x,{\bf k}_{\perp})$ ranges
from 0 to 1; thus $\Delta q$ measured
in polarized deep inelastic scattering cannot be
identified
with the spin carried by each quark flavor in the proton
rest frame \cite{Ma91b}. 

From the above discussions concerning the Wigner rotation,
we can write the quark helicity distributions
for the u and d quarks as
\begin{equation}
\begin{array}{clcr}
\Delta u_{v}(x)=u_{v}^{\uparrow}(x)-u_{v}^{\downarrow}(x)=
-\frac{1}{18}a_V(x)W_V(x)+\frac{1}{2}a_S(x)W_S(x);\\
\Delta d_{v}(x)=d_{v}^{\uparrow}(x)-d_{v}^{\downarrow}(x)
=-\frac{1}{9}a_V(x)W_V(x).
\label{eq:sfdud}
\end{array}
\end{equation}
From Eq.~(\ref{eq:ud}) one gets
\begin{equation}
\begin{array}{clcr}
a_S(x)=2u_v(x)-d_v(x);\\
a_V(x)=3d_v(x).
\label{eq:qVS}
\end{array}
\end{equation}
Combining Eqs.~(\ref{eq:sfdud}) and (\ref{eq:qVS}) we have
\begin{equation}
\begin{array}{clcr}
\Delta u_{v}(x)
    =[u_v(x)-\frac{1}{2}d_v(x)]W_S(x)-\frac{1}{6}d_v(x)W_V(x);\\
\Delta d_{v}(x)=-\frac{1}{3}d_v(x)W_V(x).
\label{eq:dud}
\end{array}
\end{equation}
Thus we arrive at simple relations \cite{Ma96} between the polarized
and unpolarized quark distributions for the valence u and d
quarks. 
The relations (\ref{eq:dud})
can be considered as the results of the conventional
SU(6) quark model 
by explicitly taking into account the Wigner rotation effect
\cite{Ma91b}
and the flavor asymmetry introduced by the
mass difference between the scalar and vector
spectators \cite{Ma96};
thus any evidence for the
invalidity of Eq.~(\ref{eq:dud}) will be useful for revealing 
new physics beyond the SU(6) quark model.

The $x$-dependent Wigner rotation factor
$W_D(x)$ has been calculated in the light-cone SU(6) quark-spectator
model \cite{Ma96} 
and an asymmetry between $W_S(x)$ 
and $W_V(x)$
was observed.
The calculated polarization asymmetries
$A_1^N=2 x g_1^N(x)/F_2^N(x)$ 
including the Wigner rotation
have been found \cite{Ma96} to be in agreement
with the experimental data, at least for $x \geq 0.1$.
A large asymmetry between $W_S(x)$ and $W_V(x)$
leads to a better fit to the data.

Thus the  $u$ and $d$ asymmetry
in the lowest valence component of the nucleon and the Wigner
rotation effect due to the internal quark transversal motions
are both important for
reproducing
the observed ratio
$F_2^n/F_2^p$ and the proton, neutron,
and deuteron polarization asymmetries,
$A_1^p$,
$A_1^n$,
$A_1^d$.
For a better understanding of
the origin of polarized sea quarks implied by
the violation of the  Ellis-Jaffe sum rule \cite{Bro88},
we still need to
consider the higher Fock states implied by the nonperturbative
meson-baryon fluctuations.
In the light-cone meson-baryon fluctuation model, the net $d$ quark
helicity of the intrinsic $q \bar q$ fluctuation is negative,
whereas the net $\bar d$ antiquark helicity is zero. Therefore the
quark/antiquark asymmetry of the $d \bar d$ pairs should be
apparent in the $d$ quark and antiquark helicity distributions.
There are now explicit measurements of the helicity distributions
for the individual $u$ and $d$ valence and sea quarks by the Spin
Muon Collaboration (SMC) \cite{NSMCN}.  The helicity distributions
for the $u$ and $d$ antiquarks are consistent with zero in agreement
with the results of the light-cone meson-baryon fluctuation model of
intrinsic $q \bar q$ pairs.
The data for the quark helicity distributions
$\Delta u_{v}(x)$ and $\Delta d_{v}(x)$ are still not 
precise enough for making detailed comparison,
but the agreement with $\Delta u_{v}(x)$ seems to be good.
It seems that 
there is some evidence for an additional source of negative
helicity contribution to the valence d quark beyond the
conventional quark model.
This again
supports the light-cone meson-baryon fluctuation model in which the
helicity distribution of the intrinsic $d$ sea quarks $\Delta
d_s(x)$ is negative.

\section{The strange quark/antiquark asymmetry in the nucleon
sea}

\begin{figure}[htb] 
\begin{center}
\mbox{\epsfysize=6cm \epsffile{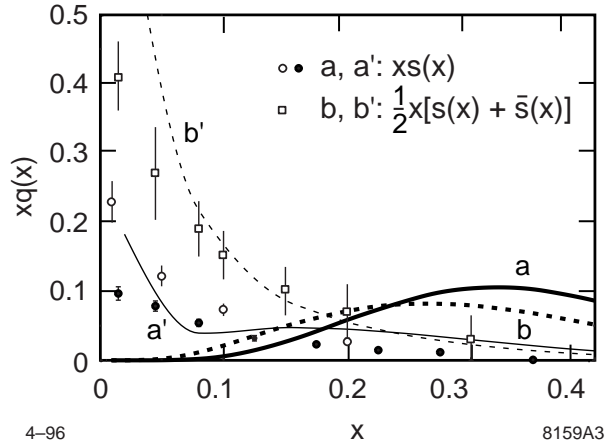}}
\end{center}
\caption[*]{\baselineskip 8pt  
Results for the strange quark distributions $x s(x)$ and $x \bar
s(x)$ as a function of the Bjorken scaling variable $x$. The open
squares shows the CTEQ determination \cite{CTEQ93} 
of $\frac{1}{2}\,x[s(x)+\bar
s(x)]$ obtained from $\frac{5}{12}\,(F_{2}^{\nu{\cal
N}}+F_{2}^{\overline{\nu}{\cal N}})(x)({\rm CCFR})- 3F_{2}^{\mu
{\cal N}}(x)({\rm NMC}).$  The circles show the CCFR determinations
for $x s(x)$ from dimuon events in neutrino scattering using a
leading-order QCD analysis at $Q^{2} \approx 5 ({\rm GeV/c})^{2} $
(closed circles) \cite{CCFR93} 
and a  higher-order QCD analysis at $Q ^2 =20 ({\rm
GeV/c})^2$ (open circles) \cite{CCFR95}. The thick curves are the unevolved
predictions of the light-cone fluctuation model for $x s(x)$ (solid
curve labeled a) and $\frac{1}{2}x[s(x)+\bar s(x)]$ (broken curve
labeled b) for the Gaussian type wavefunction in the light-cone
meson-baryon fluctuation model of intrinsic $q \bar q$ pairs
assuming a  probability of 10\% for the $K^+ \Lambda$ state. The
thin solid and broken curves (labeled a' and b') are the
corresponding evolved predictions multiplied by 
$d_v(x)|_{fit}/d_v(x)|_{model}$ assuming  a probability of 4\% for
the $K^+ \Lambda$ 
state. ~~~~ ~~~ 
}
\label{bmfig3}
\end{figure}

We have shown that the light-cone meson-baryon fluctuation model of
intrinsic $q \bar q$ pairs leads to significant quark/antiquark
asymmetries in the momentum and helicity distributions of the
nucleon sea quarks.
There is still no direct experimental confirmation of the
strange/antistrange asymmetry, although there have been
suggestions from estimates in the cloudy bag model
\cite{Sig87} and Skyrme solutions to chiral theories \cite{Bur92}.
However,
there are difficulties in understanding the
discrepancy
between two different determinations
of the strange quark content in the nucleon
sea \cite{CTEQ93,CCFR93,CCFR95}
assuming conventional considerations \cite{Ma96b} and
perturbative QCD effects \cite{Glu96}. It has been shown
that
a strange/antistrange
momentum asymmetry in the nucleon can be inferred from the apparent
strange conflict \cite{Bro96}, as can be seen from Fig.~\ref{bmfig3}.

Although the quark/antiquark asymmetries for
the intrinsic $q \bar q$ pairs can
explain the above conflict between two different determinations of the
strange quark sea in the nucleon
\cite{Bro96}, there is still no
direct experimental confirmation for such asymmetries.
If there are significant quark/antiquark asymmetries
in the distribution of the
$s \bar s$ pairs in the nucleon sea,
corresponding asymmetries should appear in
the jet fragmentation of $s$ versus $\bar s$ quarks into nucleons.
For example,
if one can identify a pure sample of tagged $s$ jets,
then one could look at the
difference of
$D_{p/s}(z)-D_{\bar p/s}(z)$ at large $z$, where $D_{h/q}(z)$ is the
fragmentation function representing the probability distribution
for the fragmentation of the
quark $q$ into hadron $h$ and $z$ is the fraction
of the quark momentum carried
by the fragmented hadron.
It has been shown \cite{Bro97}
that the hadronic jet fragmentation of
the $s$ and $c$ quarks in electron-positron ($e^+e^-$) annihilation
may provide a feasible laboratory for identifying
quark/antiquark asymmetries in the nucleon sea.

The strange quark-antiquark asymmetry implies a nonzero
strangness contribution to the magnetic moment of the nucleon.
The predictions of the meson-baryon
fluctuation model and the uncertainties due to isospin symmetry
breaking between the proton and neutron for the experimental extraction
of the strangeness contribution to nucleon moments are
discussed in ref. \cite{Ma97}.

\section{Summary}

Intrinsic sea quarks clearly play a key role in determining basic
properties of the nucleon.
As we
have shown above, the corresponding intrinsic contributions to the
sea quark structure functions lead to nontrivial, asymmetric, and
structured momentum and spin distributions. 
We have studied the intrinsic sea quarks in
the nucleon wavefunction which are generated by a light-cone model
of energetically-favored meson-baryon fluctuations.  
Such a model is supported by experimental phenomena related to the
proton spin problem: the recent SMC measurement of helicity
distributions for the individual up and down valence quarks and sea
antiquarks, and the global fit of different quark helicity contributions
from experimental data {\it et al.}
The light-cone meson-baryon fluctuation model
also suggests a structured momentum distribution asymmetry for
strange quarks and antiquarks which is related to an outstanding
conflict between two different measures of strange quark sea in the
nucleon. The model predicts an excess of intrinsic $d \bar d$ pairs
over $u \bar u$ pairs, as supported by the Gottfried sum rule
violation. We also predict that the intrinsic charm and anticharm
helicity and momentum distributions are not identical.

The intrinsic sea model thus gives a clear picture of quark flavor
and helicity distributions, which is supported qualitatively by a number of
experimental phenomena.  It seems to be an important physical source
for the violation of the Gottfried and Ellis-Jaffe sum rules and the conflict
between two different measures of strange quark distributions.

\newpage

{

}

\section*{References}
\begin{thebibliography}{99}

\bibitem{Kum97}
                For a recent review, see, e.g., 
                S.~Kumano, 
                hep-ph/9702367. 

\bibitem{SpinR} For recent reviews, see, e.g., 
H.-Y.~Cheng, Int.~J.~Mod.~Phys.~{\bf A 11} (1996) 5109; 
G.P.~Ramsey, 
hep-ph/9702227.

\bibitem{EMC82} EM Collab., J.~J.~Aubert et al, Phys.~Lett.~{\bf B 110}
(1982) 73.

\bibitem{Bro96}
S.J.~Brodsky and B.-Q.~Ma, Phys.~Lett.~{\bf B 381} (1996) 317.

\bibitem{Bro81}
S.~J.~Brodsky, P.~Hoyer, C.~Peterson, and N.~Sakai, Phys. Lett.
{\bf B 93} (1980) 451;
S.~J.~Brodsky, C.~Peterson, and N.~Sakai, Phys. Rev.
{\bf D 23} (1981) 2745.

\bibitem{Vog96} B.W. Harris, J. Smith, and R. Vogt, Nucl. Phys. {\bf B 461}
(1996) 181.

\bibitem{Ma91}
B.-Q.~Ma, Phys. Rev. {\bf C 43} (1991) 2821; Int. J. Mod. Phys. {\bf
E 1} (1992) 809.

\bibitem{Ma91b}
B.-Q.~Ma, J. Phys. {\bf G 17} (1991) L53; B.-Q.~Ma and Q.-R.~Zhang,
Z.~Phys. {\bf C 58} (1993) 479; S.~J.~Brodsky and F.~Schlumpf, Phys.
Lett. {\bf B 329} (1994) 111.

\bibitem{Ma96}
B.-Q.~Ma, Phys. Lett. {\bf B 375} (1996) 320.

\bibitem{NMC91} NM Collab., P.~Amaudruz {\it et al.},
                Phys.~Rev.~Lett. {\bf 66} (1991) 2712;
                M.~Arneodo {\it et al.},
                Phys.~Rev.~{\bf D 50} (1994) R1.

\bibitem{Pre91} G.~Preparata, P.~G.~Ratcliffe, and J.~Soffer,
                Phys.~Rev.~Lett. {\bf 66} (1991) 687.

\bibitem{Pi}    E.~M.~Henley and G.~A.~Miller,
                Phys.~Lett.~{\bf B 251} (1990) 453;
                S.~Kumano, Phys.~Rev.~{\bf D 43} (1991) 59;
                {\bf D 43} (1991) 3067;
                A.~Signal, A.~W.~Schreiber, and A.~W.~Thomas,
                Mod.~Phys.~Lett.~{\bf A 6} (1991) 271.

\bibitem{Ma92}  B.~-Q.~Ma, Phys.~Lett.~{\bf B 274} (1992) 111;
B.~-Q.~Ma, A.~Sch\"afer, and W.~Greiner,
                Phys.~Rev.~{\bf D 47} (1993) 51;
                J.~Phys.~{\bf G 20} (1994) 719.

\bibitem{Bro88} S.~J.~Brodsky, J.~Ellis, and M.~Karliner,
        Phys.~Lett.~{\bf 206 B} (1988) 309

\bibitem{NSMCN}
SM Collab., B.~Adeva {\it et al.}, Phys.~Lett.~{\bf B 369} (1996) 93.

\bibitem{Sig87}
A.~I.~Signal and A.~W.~Thomas, Phys.~Lett.~{\bf B 191} (1987) 205.

\bibitem{Bur92}
M.~Burkardt and B.~J.~Warr, Phys.~Rev.~{\bf D 45} (1992) 958.

\bibitem{CTEQ93}
CTEQ Collab., J.~Botts {\it et al.}, Phys.~Lett.~{\bf B 304} (1993)
159.

\bibitem{CCFR93}
S.~A.~Rabinowitz {\it et al.}, Phys.~Rev.~Lett. {\bf 70} (1993) 134.

\bibitem{CCFR95}
CCFR Collab., A.~O.~Bazarko {\it et al.}, Z.~Phys.~{\bf C 65} (1995)
189.

\bibitem{Ma96b} B.-Q.~Ma, Chin. Phys. Lett. {\bf 13} (1996) 648.

\bibitem{Glu96} M.~Gl\"uck, S.~Kretzer, and E.~Reya, Phys.~Lett.~{\bf B 380 }
(1996) 171. 

\bibitem{Bro97}
S.J.~Brodsky and B.-Q.~Ma,
Phys.~Lett.~{\bf B 392} (1997) 452.

\bibitem{Ma97}
B.-Q.~Ma, hep-ph/9707226,
Phys.~Lett.~{\bf B} (1997) in press.


\nonfrenchspacing
\end{thebibliography}
\end{document}